\documentclass[letterpaper, 10 pt, conference]{ieeeconf}  

\IEEEoverridecommandlockouts                              

\usepackage{cite}
\usepackage{amsmath,amssymb,amsfonts}
\usepackage{graphicx}
\usepackage{textcomp}

\usepackage[utf8]{inputenc}
\usepackage{graphics} 
\usepackage{epsfig} 
\usepackage{times} 

\usepackage{graphicx}      
\usepackage{siunitx}
\usepackage{pgfplots}
\usepackage{tikz,amsmath}
\usepackage{tikzscale}
\usepackage{algpseudocode}
\usepackage{amsthm}
\usepackage{algorithm}
\usepackage{fancyhdr}
\usepackage{subcaption} 
\usepackage{eso-pic}
\usepackage{mathdots}
\usepackage{accents}
\usetikzlibrary{calc}
\usepackage{comment}
\usepackage{mathtools}

\usetikzlibrary{shapes,arrows}
\usetikzlibrary{positioning}
\usetikzlibrary{spy}

\let\labelindent\relax
\usepackage[shortlabels]{enumitem}

\newtheorem{thm}{Theorem}

\newtheorem{prop}[thm]{Proposition}

\theoremstyle{plain}
\newtheorem{assumption}{Assumption}

\pgfplotsset{compat=newest}
\pgfplotsset{plot coordinates/math parser=false}
\newlength\figureheight
\newlength\figurewidth

\newlength\defcolwidth

\newcommand*{\tran}{^{\mkern-1.5mu\mathsf{T}}}

\usepackage{tikz}
\usepackage{tikzscale}

\definecolor{myorange}{cmyk}{0,0.35,0.85,0} 
\definecolor{mypurple}{cmyk}{0.5,1,0,0} 

\definecolor{matblue1}{rgb}{0,0.4470,0.7410}
\definecolor{matred1}{rgb}{0.85,0.325,0.098}
\definecolor{matyel1}{rgb}{0.9290, 0.6940, 0.1250}
\definecolor{matpur1}{rgb}{0.4940, 0.1840, 0.5560}
\definecolor{matgre1}{rgb}{0.4660, 0.6740, 0.1880}
\definecolor{matblue2}{rgb}{0.3010, 0.7450, 0.9330}
\definecolor{matred2}{rgb}{0.6350, 0.0780, 0.1840}
\definecolor{matgrey1}{rgb}{0.5, 0.6, 0.7}
\definecolor{matpink1}{rgb}{1, 0.07, 0.65}
\definecolor{matblue3}{rgb}{0.07, 0.62, 1}

\newcommand{\yeldots}{\raisebox{2pt}{\tikz{\draw[-,matyel1,densely dotted,line width = 0.9pt](0,0) -- (3mm,0);}}}

\newcommand{\reddash}{\raisebox{2pt}{\tikz{\draw[-,matred1,dashed,line width = 0.9pt](0,0) -- (3mm,0);}}}

\newcommand{\blueline}{\raisebox{2pt}{\tikz{\draw[-,matblue1,solid,line width = 0.9pt](0,0) -- (3mm,0);}}}
\newcommand{\redline}{\raisebox{2pt}{\tikz{\draw[-,matred1,solid,line width = 0.9pt](0,0) -- (3mm,0);}}}

\newcommand{\yelline}{\raisebox{2pt}{\tikz{\draw[-,matyel1,solid,line width = 0.9pt](0,0) -- (3mm,0);}}}

\graphicspath{{./Figures/}}

\DeclareMathOperator*{\argmax}{arg\,max}

\title{\LARGE \bf
Uncertainty-Based Perturb and Observe for Fast Optimization of Unknown, Time-Varying Processes*
}%

\author{Leontine Aarnoudse$^{1}$, Mark Haring$^{2}$, Nathan van de Wouw$^{3}$, and Alexey Pavlov$^{1}$
	\thanks{*This work was supported by the Research Council of Norway (RCN) through the project EXTREME EFFICIENCY: Data-driven optimization of industrial processes in time-varying environments (RCN project nr. 345272).}
	\thanks{$^{1}$Leontine Aarnoudse and Alexey Pavlov are with the Dept. for Geoscience, Norwegian University of Science and Technology, Trondheim, Norway. {\tt\small leontine.i.m.aarnoudse@ntnu.no}}%
	\thanks{$^{2}$Mark Haring is with the Dept. of Mathematics and Cybernetics, SINTEF Digital, Trondheim, Norway.
		}%
	\thanks{$^{3}$Nathan van de Wouw is with the Dept. of Mechanical Engineering, Eindhoven University of Technology, Eindhoven, The Netherlands.
		}%
}%

\begin{document}
	\AddToShipoutPictureBG*{%
	\AtPageUpperLeft{%
		\setlength\unitlength{1in}%
		\hspace*{\dimexpr0.5\paperwidth\relax}
		\makebox(0,-1)[c]{
			\parbox{\paperwidth}{ \centering
				Leontine Aarnoudse, Mark Haring, Nathan van de Wouw, and Alexey Pavlov, \\ Uncertainty-Based Perturb and Observe for Fast Optimization of Unknown, Time-Varying Processes, \\
				To appear in {\em Conference on Decision and Control 2025}, Rio de Janeiro, Brazil, 2025}}%
}}

\maketitle%
\thispagestyle{empty}%
\pagestyle{empty}%

\setlength\defcolwidth{7.85cm}

\setlength\figurewidth{.9\defcolwidth}
\setlength\figureheight{.7\figurewidth}

\begin{abstract}
	Model-free adaptive optimization methods are capable of optimizing unknown, time-varying processes even when other optimization methods are not. However, their practical application is often limited by perturbations that are used to gather information on the unknown cost and its gradient. The aim of this paper is to develop a perturb-and-observe (P\&O) method that reduces the need for such perturbations while still achieving fast and accurate tracking of time-varying optima. To this end, a (time-varying) model of the cost is constructed in an online fashion, taking into account the uncertainty on the measured performance cost as well as the decreasing reliability of older measurements. Perturbations are only used when this is expected to lead to improved performance over a certain time horizon. Convergence conditions are provided under which the strategy converges to a neighborhood of the optimum. Finally, simulation results demonstrate that uncertainty-based P\&O can reduce the number of perturbations significantly while still tracking a time-varying optimum accurately.
\end{abstract}

\section{Introduction} \label{sec:intro}

%
%
%

Many industrial processes are challenging to optimize, and the potential to increase, e.g., energy efficiency or other performance indicators is often not realized. Especially in sectors such as the energy and process industry, changing environmental conditions lead to time-varying processes with large uncertainties. Other contributors to this problem include uneven feed compositions, limited sensor measurements, upstream variations in the production line and a general lack of in-depth understanding of the process characteristics. Applying model-based control to this type of processes leads to conservative control and limited performance due to robustness requirements for large uncertainties in the models.

In contrast to model-based methods, model-free adaptive optimization methods such as extremum seeking control (ESC) \cite{Krstic2000,Scheinker2024} and perturb and observe (P\&O) \cite{Hussein1995} can optimize these types of processes. ESC and P\&O operate by actively probing the process to improve the operating conditions. In typical ESC, a periodic dither signal is used to perturb tunable process parameters in order to determine a direction of improvement for the parameters, for example a gradient. When determining the direction of improvement, a form of averaging is applied based on the periodic dither signal such that the influence of measurement disturbances is small. However, using a periodic dither can be inefficient and the optimization can be slow due to the required time-scale separation between the plant dynamics, the gradient estimator and the optimizer. An exception is ESC for feedback control, which uses fast perturbations but requires knowledge of a reference to track \cite{Scheinker2024}. In P\&O, the parameters are moved in one direction, and the direction is changed when the performance degrades. This leads to faster optimization, but the method is sensitive to measurement noise. 

Model-free adaptive optimization methods are promising for improving performance and energy efficiency in many applications, including wind energy systems \cite{Bafandeh2017,Mulders2019}, solar arrays \cite{Killi2015}, drilling \cite{Nystad2022}, ovens \cite{Yilmaz2023} and motion stages \cite{Hazeleger2022a}. Still, the roll-out of these solutions in industry is hampered by choices regarding the perturbations. Both ESC and P\&O continuously perturb the control parameters even after converging to the optimum, respectively through periodic dither or by a constant switching of directions around the optimum. Some parameter changes remain necessary to track the changing optimum for a time-varying process, but ESC and P\&O use more perturbations than needed, leading to reduced performance, increased wear of machinery, unnecessary use of resources, and unnecessarily large operating costs. 

Several methods have been developed to reduce these unnecessary perturbations. When P\&O is applied to maximum power point tracking, it is often attempted to look at other parameters based on the underlying physical model \cite{Hussein1995,Femia2007,Zhang2009a}. This solution, however, requires system knowledge and is limited to one specific application. In \cite{Buyukdegirmenci2010}, the step size of P\&O is adapted based on changes in the measured function value. In \cite{Moura2013,Bafandeh2017}, a Lyapunov-based switching scheme is used in ESC to estimate the system's proximity to the optimum, and the perturbation decays when closer to the optimum. Similar ideas appear in \cite{Moase2010,Atta2016,Bhattacharjee2021}, where the perturbation magnitude is scaled based on the gradient estimate, using that the gradient reduces close to the optimum. In these approaches, it may be difficult to detect a changing optimum because of the reduced perturbation magnitudes, and time-varying optima might not be tracked accurately. In the context of ESC for (stabilizing) control, it is also possible to turn of ESC if the error is small \cite{Scheinker2016}, but this requires knowledge of the output value of the optimum.

Although significant steps have been taken to improve the suitability of model-free adaptive optimization methods for industrial implementations, a method that reduces unnecessary perturbations while still achieving fast tracking of time-varying optima with limited sensitivity to noise is lacking. This research aims to develop an uncertainty-based perturb-and-observe (uP\&O) method that takes into account the uncertainty due to noisy measurements and the decreasing reliability of older data due to the time-varying nature of the process. Based on a model of the cost, the parameters are only perturbed when this is expected to improve the performance over a certain time horizon. The main contribution of the current paper consists of the following elements:
\begin{itemize}
	\item A new uP\&O method is presented (Section \ref{sec:approach}).
	\item Convergence conditions for P\&O and uP\&O are developed (Section \ref{sec:convergence}).
	\item Simulations of a photovoltaic array to illustrate that uP\&O requires far fewer perturbations than P\&O to accurately track a time-varying optimum (Section \ref{sec:sims}).
\end{itemize} 

\noindent
\textit{Notation:} $\mathbb{R} $ and $\mathbb{N}$ denote respectively the sets of real and natural numbers. Subscripts denote discrete time instants (e.g., $u_k$ denotes the value of $u$ at time instant $k$), while superscripts in parentheses denote a member of a discrete set (e.g., $u^{(i)}$ denotes the $i$-th element of a set $\{u^{(1)},u^{(2)},...\}$).

\section{Problem formulation} \label{sec:problem}

\begin{figure}[t]
	\centering
	\setlength\figureheight{.35\figurewidth}
	\includegraphics{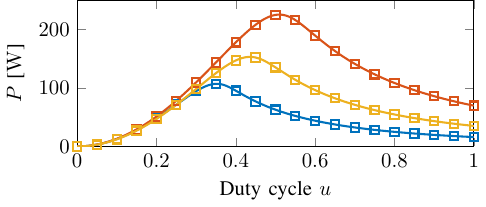}
	\caption{The aim is to find and track the optimum of a time-varying function $f_k(u)$, e.g., the steady-state produced power $P$ of a photovoltaic array as a function of the duty cycle $u$, shown at times $k=50$ (\protect\blueline), $150$ (\protect\redline) and $250$ (\protect\yelline), see also Section \ref{sec:sims}. The squares indicate the possible inputs. \label{fig:graph}}
	\vspace{-15pt}
\end{figure}

Consider the problem of optimizing an unknown, static, time-varying function $f_k(u_k)$, with input values $u_k \in \mathcal{U} \subset \mathbb{R}$ for time index $k \in \mathbb{N}$, as illustrated in Fig. \ref{fig:graph}. The set of equidistantly spaced input values at which the function can be measured is given by $\mathcal{U} = \{u^{(1)}, u^{(2)},...,u^{(N_u)}\}$ for $N_u \in \mathbb{N}_{>0}$. The function measurements $y_k$ are given by
\begin{align} \label{eq:yk}
	y_k = f_k (u_k) + \rho \varepsilon_k,
\end{align}
with $\rho \in \mathbb{R}_{>0}$, and $\varepsilon_k \in \mathbb{R}$ a white, standard Gaussian variable, that captures the error between the function value and the measurement. The following assumption is adopted.

\begin{assumption} \label{ass:min}
	At each time $k$, there exists a unique optimizer $u_k^* = \argmax_{u\in \mathcal{U}} f_k(u)$.
\end{assumption}

A common approach to find and track the maximum $u_k^*$ is perturb and observe. The input is perturbed to determine the direction of improvement $g_k \in\{-1,1\} $, which follows from two subsequent output measurements as
\begin{align} g_{k+1}=
	\begin{cases}
		g_k, & \text{if }  y_k\geq y_{k-1}, \\
		-g_k & \text{if } y_k < y_{k-1}.
	\end{cases} \label{eq:gk}
\end{align}
This formula is used to update the input according to
\begin{align} \label{eq:uk}
	u_{k+1} = u_k + g_{k+1} \Delta_u,
\end{align}
where $\Delta_u = u^{(j+1)}-u^{(j)}$ is the distance between any two neighboring inputs. Thus, the direction in which the input is updated remains constant if the function value increases, and if the function value reduces the direction is reversed.

P\&O can find and track optima effectively, but it has two main disadvantages. First, it is sensitive to noise, which can cause the algorithm to move in the wrong direction. Second, after the optimum is found, the algorithm continues perturbing the input. Although perturbations are needed to track the changing optimum of the time-varying function, it has been observed in a number of examples that these continuing, unconditional perturbations are excessive and may reduce the overall performance or lead to increased wear and tear of the system. This paper aims to develop a perturbation-based optimization method that can find and track the optimum even in the presence of large disturbances, while significantly reducing the number of perturbations.

\section{Uncertainty-based perturb and observe} \label{sec:approach}

In this section, a new \textit{uncertainty-based} perturb-and-observe algorithm is introduced that reduces the number of perturbations needed to find and track time-varying optima. First, a model for the time-varying function is constructed online from current and older measurements, while taking into account measurement uncertainty and variation over time. Second, this model is used to find the optimal input over a horizon. Third, implementation aspects are considered.

\subsection{Uncertainty-based model for the cost} \label{subsec:model}

Consider the function $f_k(u_k)$ with measurements according to \eqref{eq:yk}. Since the function changes over time, older measurements give potentially outdated and uncertain information about the current function value at the same input. To estimate current function values, we model the relation between older measurements and current function values using an uncertainty that increases over time:
\begin{align} \label{eq:measurment_model}
	y_j = f_k(u) + \textstyle\frac{\rho}{\lambda^{k-j}} \varepsilon_j, \quad \mbox{if $u_j = u$},
\end{align}
for forgetting factor $\lambda \in (0,1]$, $u \in \mathcal{U}$ and all $k,j \in \mathbb{N}$ such that $k \geq j$. The forgetting factor $\lambda \in (0,1]$ should be small for functions that change quickly over time, and close to 1 for functions that change slowly. We capture all available information about the function value at $u$ in the model:
{\small \begin{equation} \label{eq:model}
	\mathcal{\widehat{M}}_{k+1}(u) = \left\{ y_j = \tilde{f}_{k+1}(u) + \textstyle\frac{\rho}{\lambda^{k+1-j}} \varepsilon_j, \, \forall j \in \mathcal{J}_{k}(u) \right\},
\end{equation}} 

\noindent where the model variable $\tilde{f}_{k+1}(u) \in \mathbb{R}$ represents the unknown function value $f_{k+1}(u)$. The index set $\mathcal{J}_k(u)$ contains all time indices up to time $k$ for which $u_j = u$ as follows
\begin{equation}
	\mathcal{J}_k(u) = \left\{j \in \{0,1,\dots,k\} : \, u_j = u \right\},
\end{equation}
for all $k\in \mathbb{N}$ and $u \in \mathcal{U}$. Solving the model \eqref{eq:model} for $\tilde{f}_{k+1}(u)$ leads to the function value estimate $\hat{f}_{k+1}(u)$, given by
{\small \begin{align} \label{eq:function_next}
	\hat{f}_{k+1}(u) = \frac{\sum_{j \in \mathcal{J}_k(u)} \lambda^{2(k-j)} y_j}{\sum_{j \in \mathcal{J}_k(u)} \lambda^{2(k-j)} } - \frac{\rho}{\lambda} \frac{\sum_{j \in \mathcal{J}_k(u)} \lambda^{k-j} \varepsilon_j}{\sum_{j \in \mathcal{J}_k(u)} \lambda^{2(k-j)} }.
\end{align}}

\noindent Since $\hat{f}_{k+1}(u)$ is a linear function of Gaussian variables, it is also Gaussian, with mean $\hat{\mu}^f_{k+1}(u)$ and variance $\hat{\sigma}^f_{k+1}(u)$:
\begin{align} \label{eq:mean_next}
	\hat{\mu}^f_{k+1}(u) &= \frac{\sum_{j \in \mathcal{J}_k(u)} \lambda^{2(k-j)} y_j}{\sum_{j \in \mathcal{J}_k(u)} \lambda^{2(k-j)} } \\ 
 \label{eq:variance_next}
	\hat{\sigma}^f_{k+1}(u) &=  \frac{1}{\lambda^2} \frac{\rho^2}{\sum_{j \in \mathcal{J}_k(u)} \lambda^{2(k-j)}}.
\end{align}
Similar to the one-time-step-ahead prediction $\hat{f}_{k+1}(u)$ in \eqref{eq:function_next}, the estimate for the current function value is given by
{\small \begin{align} \label{eq:function_now}
	\bar{f}_{k}(u) = \frac{\sum_{j \in \mathcal{J}_k(u)} \lambda^{2(k-j)} y_j}{\sum_{j \in \mathcal{J}_k(u)} \lambda^{2(k-j)} } - \rho \frac{\sum_{j \in \mathcal{J}_k(u^{(i)})} \lambda^{k-j} \varepsilon_j}{\sum_{j \in \mathcal{J}_k(u)} \lambda^{2(k-j)} }.
\end{align}}

\noindent The mean $\bar{\mu}^f_{k}(u)$ and variance $\bar{\sigma}^f_{k}(u)$ of current function value estimate $\bar{f}_k(u)$ relate to the mean and variance of the one-time-step-ahead prediction $\hat{f}_{k+1}(u)$ according to
\begin{align} \label{eq:mean_next_now}
	\hat{\mu}^f_{k+1}(u) &= \bar{\mu}^f_{k}(u) , \quad 
	\hat{\sigma}^f_{k+1}(u) = \textstyle\frac{1}{\lambda^2} \bar{\sigma}^f_{k}(u).
\end{align}
Since $\hat{f}_{k+1}(u)$ is Gaussian, it can be written in the following equivalent form, with standard Gaussian variable $\varepsilon^f_{k+1} \in \mathbb{R}$:
\begin{align} \label{eq:function_next_equivalent}
	\hat{f}_{k+1}(u) = \hat{\mu}^f_{k+1}(u) + \sqrt{\hat{\sigma}^f_{k+1}(u)} \varepsilon^f_{k+1}.
\end{align}
Using \eqref{eq:mean_next_now} and \eqref{eq:function_next_equivalent}, it follows that
\begin{align} \label{eq:function_approximation}
	\hat{f}_{k+1}(u; \mathbf{s}_k) = \bar{\mu}^f_{k}(u) + \textstyle\frac{1}{\lambda} \sqrt{\bar{\sigma}^f_{k}(u)} \varepsilon^f_{k+1},
\end{align}
with the model state $\mathbf{s}_k$, which contains the mean and variance of the estimates of all function values, given by
{\small \begin{align} \label{eq:state} \setlength\arraycolsep{2pt}
	\mathbf{s}_k = \begin{bmatrix} \bar{\mu}^f_{k}(u^{(1)}) & \dots & \bar{\mu}^f_{k}(u^{(N)}) & \bar{\sigma}^f_{k}(u^{(1)}) & \dots & \bar{\sigma}^f_{k}(u^{(N)})\end{bmatrix}\tran.
\end{align}}
\noindent Next, this model state is used to determine the next input.

\subsection{Input selection} \label{subsec:input}

The best next input value $u_{k+1}$ is determined by considering its expected influence on the future function values. The aim is to minimize the function value over an evaluation horizon consisting of $p \in \mathbb{N}_{>0}$ future time steps. For any given model state, we define the best next input value as 
\begin{align} \label{eq:ustar}
	u^*_{k+1}(\mathbf{s}_k) =& \argmax_{u_{k+1} \in \mathcal{U}} \mathbb{E}\left\{ \hat{f}_{k+1}(u_{k+1}; \mathbf{s}_{k})  + \dots \right. \\ \nonumber  &+ \max_{u_{k+r} \in \mathcal{U}} \mathbb{E}\left\{ \hat{f}_{k+r}(u_{k+r}; \mathbf{s}_{k+r-1})  \right. \\ \nonumber & \left. \left. + \max_{u_{k+p} \in \mathcal{U}} \mathbb{E}\left\{ \hat{f}_{k+p}(u_{k+p}; \mathbf{s}_{k+p-1}) \right\} \right\} \right\},
\end{align}
for all $r\in \{2,3,\dots,p-1\}$. For any $p \geq 2$, the optimal input value can be written as
\begin{align} \nonumber 
	&u_{k+1}^*(\mathbf{s}_k) = \argmax_{u_{k+1} \in \mathcal{U}} \mathbb{E}\left\{ \hat{f}_{k+1}(u_{k+1}; \mathbf{s}_{k}) + V_{k,1}(\mathbf{s}_{k+1}) \right\}, \\ \nonumber 
&	V_{k,r}(\mathbf{s}_{k+r}) =  \max_{u_{k+r+1} \in \mathcal{U}} \mathbb{E}\left\{ \hat{f}_{k+r+1}(u_{k+r+1}; \mathbf{s}_{k+r}) \right. \\ \nonumber &\left. \qquad + V_{k,r+1}(\mathbf{s}_{k+r+1}) \right\}, \: \forall r \in \{1,2,\dots,p-2\}, \: \text{and} \\  \label{eq:optimal_input}
&	V_{k,p-1}(\mathbf{s}_{k+p-1}) = \max_{u_{k+p} \in \mathcal{U}} \mathbb{E}\left\{ \hat{f}_{k+p}(u_{k+p}; \mathbf{s}_{k+p-1}) \right\}.
\end{align}
Note that the future states $\mathbf{s}_{k+1}$ up to $\mathbf{s}_{k+p-1}$ in \eqref{eq:optimal_input} are not explicitly known. Instead, a stochastic approximation of the future state vectors $\mathbf{s}_{k+r+1}$ with means $\bar{\mu}^f_{k+r+1}(u)$ and variances $\bar{\sigma}^f_{k+r+1}(u)$ for all $u \in \mathcal{U}$ is computed. It holds that
\begin{align} \label{eq:mean_now_next}
	\bar{\mu}^f_{k+r+1}(u) &= \frac{\sum_{j \in \mathcal{J}_{k+r+1}(u)} \lambda^{2(k+r+1-j)} y_j}{\sum_{j \in \mathcal{J}_{k+r+1}(u)} \lambda^{2(k+r+1-j)} }, \\
 \label{eq:variance_now_next}
	\bar{\sigma}^f_{k+r+1}(u) &=  \frac{\rho^2}{\sum_{j \in \mathcal{J}_{k+r+1}(u)} \lambda^{2(k+r+1-j)}}.
\end{align}

This can be rewritten to
{\small\begin{align} \label{eq:next_mu}
	&\bar{\mu}^f_{k+r+1}(u) = \\ \nonumber & \begin{cases}
		\bar{\mu}^f_{k+r}(u) + K_{k+r}(u) (y_{k+r+1} - \bar{\mu}^f_{k+r}(u)), & \mbox{if $u_{k+r+1} = u$,}\\
		\bar{\mu}^f_{k+r}(u), & \mbox{if $u_{k+r+1} \neq u$.}
	\end{cases}
\\  \nonumber
	&\bar{\sigma}^f_{k+r+1}(u) = \begin{cases}
		 \left(1 - K_{k+r}(u) \right)^2 \frac{1}{\lambda^2}\bar{\sigma}^f_{k+r}(u^{(i)}) \\ \qquad + \rho^2 \left( K_{k+r}(u^{(i)}) \right)^2,   &  \mbox{if $u_{k+r+1} = u$,}\\
		\frac{1}{\lambda^2}\bar{\sigma}^f_{k+r}(u),   &  \mbox{if $u_{k+r+1} \neq u$,}
	\end{cases}
\end{align}}

\noindent where the gain $K_{k+r}(u)$ is given by
{\small \begin{align}
	K_{k+r}(u) &= \frac{1}{1 + \lambda^2 \sum_{j \in \mathcal{J}_{k+r}(u)} \lambda^{2(k+r-j)}} = \frac{\frac{1}{\lambda^2}\bar{\sigma}^f_{k+r}(u)}{\frac{1}{\lambda^2}\bar{\sigma}^f_{k+r}(u) + \rho^2}.
\end{align}}

\noindent The measurement $y_{k+r+1}$, which is needed in \eqref{eq:next_mu}, is approximated using the model in Section \ref{subsec:model}. Using \eqref{eq:measurment_model} with the approximation $\hat{f}_{k+r+1}(u_{k+r+1};\mathbf{s}_{k+r})$ in \eqref{eq:function_approximation} gives
\begin{align}
		&\hat{y}_{k+r+1} = \hat{f}_{k+r+1}(u_{k+r+1};\mathbf{s}_{k+r}) + \rho \varepsilon_{k+r+1} \\ \nonumber
		&= \bar{\mu}^f_{k+r}(u_{k+r+1}) + \frac{1}{\lambda} \sqrt{\bar{\sigma}^f_{k+r}(u_{k+r+1})} \varepsilon^f_{k+r+1} + \rho \varepsilon_{k+r+1}. 
\end{align}
This approximation can be written as
{\small\begin{align} \label{eq:measurement_approximation}
	\hat{y}_{k+r+1} = \hat{\mu}^y_{k+r+1}(u_{k+r+1}) + \sqrt{\hat{\sigma}^y_{k+r+1}(u_{k+r+1})} \varepsilon^y_{k+r+1},
\end{align}}

\noindent with $\varepsilon^y_{k+r+1}$ a standard, Gaussian variable and
\begin{align}
	\hat{\mu}^y_{k+r+1}(u_{k+r+1}) &= \bar{\mu}^f_{k+r}(u_{k+r+1}), \\
	\hat{\sigma}^y_{k+r+1}(u_{k+r+1}) &= \frac{1}{\lambda^2} \bar{\sigma}^f_{k+r}(u_{k+r+1}) + \rho^2,
\end{align}
Substituting the measurement approximation \eqref{eq:measurement_approximation} in \eqref{eq:next_mu}, and using \eqref{eq:state} and \eqref{eq:next_mu}, leads to the following approximation of the future state $\mathbf{s}_{k+r+1}$ for all $r \in \{0,1,\dots,p-2\}$:
{\small \begin{align} \label{eq:state_approx} 
	&\mathbf{\hat{s}}_{k+r+1}(u_{k+r+1};\mathbf{s}_{k+r},\varepsilon^y_{k+r+1}) = \\ \nonumber &  \left[ \hspace{-5pt} \begin{array}{l}
		\begin{cases}
			\bar{\mu}^f_{k+r}(u^{(1)}) + K_{k+r}(u^{(1)}) \sqrt{\frac{1}{\lambda^2} \bar{\sigma}^f_{k+r}(u^{(1)}) + \rho^2} \varepsilon^y_{k+r+1}, \\ \qquad \qquad \qquad \quad \mbox{if $u_{k+r+1} = u^{(1)}$}\\
			\bar{\mu}^f_{k+r}(u^{(1)}),   \qquad \mbox{if $u_{k+r+1} \neq u^{(1)}$}
		\end{cases} \\
		\qquad \qquad \qquad \qquad \qquad \qquad \vdots \\
		 \begin{cases}
			\bar{\mu}^f_{k+r}(u^{(N)}) + K_{k+r}(u^{(N)}) \sqrt{\frac{1}{\lambda^2} \bar{\sigma}^f_{k+r}(u^{(N)}) + \rho^2} \varepsilon^y_{k+r+1}, \\ \qquad \qquad \qquad \quad \: \:  \mbox{if $u_{k+r+1} = u^{(N)}$}\\
			\bar{\mu}^f_{k+r}(u^{(N)}),  \qquad \mbox{if $u_{k+r+1} \neq u^{(N)}$}
		\end{cases}   \\
		\begin{cases}
			\left(1 - K_{k+r}(u^{(1)}) \right)^2 \frac{1}{\lambda^2}\bar{\sigma}^f_{k+r}(u^{(1)}) + \rho^2 \left( K_{k+r}(u^{(1)}) \right)^2, \\ \qquad \qquad \qquad \quad \: \: \mbox{if $u_{k+r+1} = u^{(1)}$}\\
			\frac{1}{\lambda^2}\bar{\sigma}^f_{k+r}(u^{(1)}),  \quad \mbox{if $u_{k+r+1} \neq u^{(1)}$}
		\end{cases}  \\
		\qquad \qquad \qquad \qquad \qquad \qquad  \vdots \\
		\begin{cases}
			\left(1 - K_{k+r}(u^{(N)}) \right)^2 \frac{1}{\lambda^2}\bar{\sigma}^f_{k+r}(u^{(N)}) + \rho^2 \left( K_{k+r}(u^{(N)}) \right)^2, \\ \qquad \qquad \qquad \quad \: \: \: \mbox{if $u_{k+r+1} = u^{(N)}$}\\
			\frac{1}{\lambda^2}\bar{\sigma}^f_{k+r}(u^{(N)}),  \quad \mbox{if $u_{k+r+1} \neq u^{(N)}$}
		\end{cases} 
	\end{array} \hspace{-15pt} \right]
\end{align}}

\noindent The expected values in \eqref{eq:optimal_input} can be computed using the approximations of the future states in \eqref{eq:state_approx}, leading to the following expression for the optimal next input $u^*_{k+1}$:
\begin{align} \label{eq:optimal_input_rewritten}
	&u_{k+1}^*(\mathbf{s}_k) = \argmax_{u_{k+1} \in \mathcal{U}} \left\{ \bar{\mu}^f_{k}(u_{k+1}) + \right. \\ \nonumber &\left. \textstyle\int_{-\infty}^\infty \textstyle\frac{1}{\sqrt{2 \pi}}  V_{k,1}(\mathbf{\hat{s}}_{k+1}(u_{k+1};\mathbf{s}_{k},\varepsilon^y_{k+1})) e^{-\frac{1}{2} \left(\varepsilon^y_{k+1}\right)^2} d \varepsilon^y_{k+1} \right\},
\end{align}
with
\begin{align} \label{eq:optimal_recursion_rewritten}
	V_{k,r}&(\mathbf{s}_{k+r}) = \max_{u_{k+r+1} \in \mathcal{U}} \left\{ \bar{\mu}^f_{k+r}(u_{k+r+1}) + \right. \\ \nonumber &\left. \textstyle\int_{-\infty}^\infty \textstyle\frac{1}{\sqrt{2 \pi}}   V_{k,r+1}(\mathbf{\hat{s}}_{k+r+1}(u_{k+r+1};\mathbf{s}_{k+r},\varepsilon^y_{k+r+1})) \right. \\ \nonumber &\left. \qquad e^{-\frac{1}{2} \left(\varepsilon^y_{k+r+1}\right)^2} d \varepsilon^y_{k+r+1} \right\}, \: \forall r \in \{1,2,\dots,p-2\}
 \\ \label{eq:optimal_end_rewritten}
	V_{k,p-1}&(\mathbf{s}_{k+p-1}) = \max_{u_{k+p} \in \mathcal{U}} \bar{\mu}^f_{k+p-1}(u_{k+p}).
\end{align}

The functions $V_{k,r}(\mathbf{s}_{k+r})$ in \eqref{eq:optimal_recursion_rewritten} and \eqref{eq:optimal_end_rewritten} can be complex for large horizons, such that computing the integrals in \eqref{eq:optimal_input_rewritten} and \eqref{eq:optimal_recursion_rewritten} is challenging. Instead, we approximate the integrals with finite weighted sums of function evaluations, using
\begin{align} \label{eq:cost_approx}
	&\textstyle\int_{-\infty}^\infty \textstyle\frac{1}{\sqrt{2 \pi}}  V_{k,r+1}(\mathbf{\hat{s}}_{k+r+1}(u_{k+r+1};\mathbf{s}_{k+r},\varepsilon^y_{k+r+1})) \\ \nonumber & \qquad e^{-\frac{1}{2} \left(\varepsilon^y_{k+r+1}\right)^2} d \varepsilon^y_{k+r+1} \\ \nonumber &\approx \textstyle\sum_{i = 0}^M w_i V_{k,r+1}(\mathbf{\hat{s}}_{k+r+1}(u_{k+r+1};\mathbf{s}_{k+r},v_i))
\end{align}
for $r \in \{0,1,\dots,p-2\}$. Here, $M \in \mathbb{N}_{>0}$ is the number of summed terms. The weights $w_i \in \mathbb{R}_{>0}$ and evaluation points $v_i \in \mathbb{R}$ are selected using Gaussian quadrature.

\subsection{Overview and implementation of uncertainty-based P\&O} \label{subsec:implementation}

For the practical implementation and tuning of uP\&O, it is useful to be able to recover standard P\&O behavior from the algorithm. Therefore, the approximated cost function in \eqref{eq:cost_approx} is extended by a part that enables enforcing perturbations:
\begin{align} \label{eq:cost2}
	&u_{k+1}^*(\mathbf{s}_k) \approx \argmax_{u_{k+1} \in \mathcal{U}} \left\{ 	W_{u_{k+1}} +  \bar{\mu}^f_{k}(u_{k+1}) - \right. \\ \nonumber &\left. \textstyle\sum_{i = 0}^M w_i V_{k,1}(\mathbf{\hat{s}}_{k+1}(u_{k+1};\mathbf{s}_{k},v_i)) \right\},
\end{align}
with
\begin{align}
	W_{u_{k+1}} = \begin{cases}
		0, & \text{if } u_{k+1} = u_k + g_k  \Delta_u\\
		W, & \text{otherwise}.
	\end{cases}
\end{align}
The direction $g_k$ is identical to that in P\&O in \eqref{eq:gk}. For large $W$, the input selected by \eqref{eq:cost_approx} is only preferred over the P\&O input if the difference between the expected costs is large. 

The uP\&O approach is summarized in Algorithm \ref{alg:upo}. In addition to the main method of determining the optimal input based on \eqref{eq:cost2}, two rules are included. First, if the expected value $\hat{\mu}^f_{k}(u_k)$ of the most recently measured point is smaller than that of the previously measured point $\hat{\mu}^f_{k}(u_{k-1})$, the next input is chosen as $u_{k+1} = u_{k-1}$. Second,  if the next point in the current direction of movement $u_{k} - u_{k-1}$ has not been measured yet, i.e., there is no estimate $\hat{\mu}^f_{k}(u_k + (u_k-u_{k-1}))$, the next input is chosen as $u_{k+1} = u_k + (u_{k} - u_{k-1})$. Note that in practice, the measurement uncertainty $\rho$ may need to be replaced by an estimate $\hat{\rho}$.

\begin{algorithm}[H]
	\caption{Uncertainty-based perturb and observe} 	\label{alg:upo}
	\begin{algorithmic}[1]
		\State{Initialize $u_1$ and $u_2 = u_1 + \Delta_u$, and measure $y_1$}
		\State{\textbf{for} $k=2:n_{\text{iteration}}-1$}
		\State{\quad Measure $y_k = f_k(u_k)+ \rho \varepsilon_k$ and compute $s_k$}
		\State{\quad \textbf{if} $\hat{\mu}^f_{k}(u_k) \leq \hat{\mu}^f_{k}(u_{k-1})$}
		\State{\qquad Return to previous point: $u_{k+1} = u_{k-1}$}
		\State{\quad \textbf{elseif} $u_{k}+(u_k-u_{k-1})$ has not been measured}
		\State{\qquad Choose $u_{k+1} = u_k + (u_k-u_{k-1})$}		
		\State{\quad \textbf{else}}
		\State{\qquad Choose $u_{k+1} = u^*_{k+1}(\mathbf{s}_k)$ according to \eqref{eq:cost2}}
		\State{\quad\textbf{end}}		
		\State{\textbf{end}}	
	\end{algorithmic}
\end{algorithm}

\section{Convergence of uncertainty-based P\&O} \label{sec:convergence}

In this section, convergence conditions for uP\&O are presented. First, P\&O is considered and second, these results are used to analyze the convergence properties of uP\&O.

\subsection{Convergence of standard perturb and observe}

Consider P\&O as presented in Section \ref{sec:problem} for an unknown function $f_k(u_k)$. To analyze the convergence properties, we assume that the measurement uncertainty is bounded: 

\begin{assumption} \label{ass:varepsilon}
	Variable $\varepsilon_k$ is bounded, i.e., $|\varepsilon_j|\leq 1 \: \forall j$. 
\end{assumption}

\noindent Here, $\varepsilon_k$ could, for example, have a truncated Gaussian distribution. In addition, the following assumptions are adopted.

\begin{assumption} \label{ass:lb}
	The rate of change of $f_k(u)$ between two neighboring inputs is bounded, i.e., there exist positive constants $L_b\in \mathbb{R}_{>0}$ and $L_u\in \mathbb{R}_{>0}$, $L_b \leq L_u$ such that
	\begin{align}
		& \textstyle\frac{d_{u,k}}{\Delta_u} L_b \leq |f_k(u^{j+1}) - f_k(u^{j})| \leq  \textstyle\frac{d_{u,k}}{\Delta_u} L_u,
	\end{align}	
	with $ d_{u,k} = \max\left(|u_k^*-u^j|,|u_k^*-u^{j+1}|\right)$ . 
\end{assumption}

\begin{assumption} \label{ass:lk}
	The rate of change of $f_k(u)$ between two subsequent time steps, for constant $u$, is upper bounded, i.e., there exists a positive constant $L_k \in \mathbb{R}_{>0}$ such that
	\begin{align}
		|f_{k+1}(u) - f_k(u)| \leq L_k, \: \forall \: k \in \mathbb{N}, \: u \in \mathcal{U}.
	\end{align}
\end{assumption}

\noindent Assumptions \ref{ass:varepsilon}-\ref{ass:lk} lead to the following theorem regarding the convergence of standard perturb and observe.

\begin{thm} \label{thm:conv1}
	Consider the time-varying function $f_k(u),\: u \in \mathcal{U}$ with measurements \eqref{eq:yk} and input update \eqref{eq:uk}, which satisfies Assumptions \ref{ass:min}-\ref{ass:lk}. Then, the sequence of inputs converges to and remains in the set $[u^*_k - \beta \Delta_u, u^*_k + \beta \Delta_u]$, with 
	\begin{align} \label{eq:beta}
		\beta = \textstyle\frac{L_k + 2\rho}{L_b}+1.
	\end{align}
\end{thm}

\begin{proof}
	To determine the direction of improvement at time $k+1$, measurements of $f_{k}(u_{k})$ and $f_{k+1}(u_{k+1})$ are compared. First, consider the case where $f_{k+1}(u_{k}) < f_{k+1}(u_{k+1})$.  Using Assumption \ref{ass:varepsilon} and \ref{ass:lk}, the highest possible measurement $\bar{y}_k$ at time step $k$, and the lowest possible measurement $\underline{y}_{k+1}$ at time step $k+1$, are given by
	\begin{align}
		\bar{y}_k &= \bar{f}_{k}(u_{k}) + \rho = f_{k+1}(u_{k}) + L_k + \rho. \\
		\underline{y}_{k+1} &= f_{k+1}(u_{k+1}) - \rho.
	\end{align}
	 The correct direction of improvement is obtained if $y_{k} - y_{k-1} < 0$, which is guaranteed if $\bar{y}_k - \underline{y}_{k+1} < 0$. Using that $f_{k+1}(u_{k+1}) - f_{k+1}(u_{k}) \leq - \frac{d_{u,k}}{\Delta_u} L_b$ by Assumption \ref{ass:lb} gives
	\begin{align}
		\bar{y}_k - \underline{y}_{k+1} &= f_{k+1}(u_{k}) + L_k + \rho - f_{k+1}(u_{k+1}) + \rho \\ \nonumber
		&= f_{k+1}(u_{k}) - f_{k+1}(u_{k+1}) + L_k + 2\rho \\ \nonumber
		& \leq  -\textstyle\frac{d_{u,k}}{\Delta_u}L_b + L_k + 2\rho.
	\end{align}
	Thus, in this case, $\bar{y}_k - \underline{y}_{k+1} < 0$ if $\frac{d_{u,k}}{\Delta_u} > \frac{L_k + 2\rho}{ L_b}$, which implies that if $u_k$ is at least $\beta = \frac{L_k + 2\rho}{ L_b} + 1$ steps removed from the optimum, the correct direction of improvement is obtained and the algorithm will move towards the optimum.
	
	Second, when $f_{k+1}(u_{k}) > f_{k+1}(u_{k+1})$, the correct direction of improvement is obtained if $\underline{y}_k - \bar{y}_{k+1} > 0$, with
	\begin{align}
		\underline{y}_k &= \underline{f}_{k}(u_{k}) - \rho = f_{k+1}(u_{k}) - L_k - \rho \\ 
		\bar{y}_{k+1} &= f_{k+1}(u_{k+1}) + \rho.
	\end{align}
 	Using that $f_{k+1}(u_{k}) - f_{k+1}(u_{k+1}) \geq  \frac{d_{u,k}}{\Delta_u}L_b$ gives 
	\begin{align}
		\underline{y}_k - \bar{y}_{k+1} &= f_{k+1}(u_{k}) - L_k - 2\rho -  f_{k+1}(u_{k+1}) \\ \nonumber
		&\geq \textstyle\frac{d_{u,k}}{\Delta_u}L_b - L_k - 2\rho.
	\end{align} 
	Thus, if $u_k$ is $\beta = \frac{L_k + 2\rho}{ L_b} + 1$ steps removed from the minimum $u_k^*$, the correct direction of improvement for $k+1$ is found. Therefore, the input converges to the set $[u^*_k - \beta \Delta u, u^*_k + \beta \Delta u]$ with $\beta = \frac{L_k + 2\rho}{ L_b} + 1$ and remains there.
\end{proof}

\subsection{Convergence of uncertainty-based perturb and observe}

The convergence of uP\&O depends on whether the algorithm perturbs `often enough' to track the varying optimum. In contrast to P\&O, which always perturbs, uP\&O only perturbs if this is expected to be beneficial over the time horizon. The number of perturbations depends on the horizon, the estimates of the measurement uncertainty $\rho$ and the forgetting factor $\lambda$, and the actual measurement disturbances. To analyze the convergence of uP\&O, the cost function in \eqref{eq:cost2} is considered, leading to the following proposition.

\begin{prop} \label{lem:conv_upo}
	Consider function $f_k(u), \: u \in \mathcal{U}$ with measurements \eqref{eq:yk} and input updates according to Algorithm 1, which satisfies Assumptions \ref{ass:min}-\ref{ass:lk}. For $W\to \infty$ and $\lambda \to 0$, the sequence of inputs converges to the set $[u^*_k - \beta \Delta u, u^*_k + \beta \Delta u]$, with $\beta$ according to \eqref{eq:beta}.
\end{prop}

\begin{proof}
	For $W\to \infty$ and $\lambda \to 0$, standard P\&O is recovered. In particular, taking $\lambda\to 0$  ensures that the estimates $\hat{\mu}^y_k(u_k)$ and $\hat{\mu}^y_k(u_{k-1})$ in Algorithm \ref{alg:upo} reduce to respectively $y_{k}$ and $y_{k-1}$. The convergence then follows from Theorem \ref{thm:conv1}. 
\end{proof}

Proposition \ref{lem:conv_upo} provides a starting point for the tuning of uP\&O, since the weight $W$ can be used to force perturbations and ensure convergence. Simulation results show that with good estimates of the measurement uncertainty $\rho$ and the forgetting factor $\lambda$, uP\&O is able to reduce the perturbations compared to standard P\&O while still ensuring fast tracking of changing minima. This is further illustrated in Section \ref{sec:sims}.

\section{Illustrative example} \label{sec:sims}

In this section uP\&O is applied to a case study of the optimization of a photo-voltaic solar array. First, the setup is introduced, and second, simulation results are given.

\subsection{Problem setup}

\begin{table}[t]
	\centering
	\caption{Parameters for the photovoltaic model \label{tab:pars}}
	\begin{tabular}{|l|l|} \hline
		Parameter & Description \\ \hline
		$T_r =$ \SI{298.15}{\kelvin} & Reference temperature \\	
		$I_s =$ \SI{5.61}{\ampere} & Reference short-circuit current at $T_r$ \\
		$I_0 =$ \SI{1.13e-6}{\ampere} & Nominal reverse saturation current at $T_r$ \\	
		$k_i =$ \SI{1.96e-3}{\ampere/ \kelvin} & Short-circuit current temperature coefficient \\
		$N =$ 1.81 & Ideality factor \\
		$E_g =$ \SI{1.16}{\electronvolt} & Band gap energy for silicon \\
		$k =$ \SI{1.38e-23}{\joule / \kelvin} & Boltzmann constant \\
		$q =$ \SI{1.60e-19}{\coulomb} & Charge of an electron \\
		$n_s =$ 72 & Number of photo-voltaic cells in series \\
		$R_s =$ \SI{2.83e-3}{\ohm} & Series resistance \\
		$R_p =$ \SI{8.7}{\ohm} & Parallel resistance \\ 
		$C_c =$ \SI{1}{\milli \farad} & Converter's capacitance \\
		$L_c =$ \SI{5}{\milli \henry} & Converter's inductance \\
		$R_c =$ \SI{2}{\ohm} & Converter's resistance \\		
		\hline
	\end{tabular}
\end{table}

\begin{figure}[t]
	\centering
	\setlength\figurewidth{.8\defcolwidth}
	\setlength\figureheight{.25\defcolwidth}
	\includegraphics{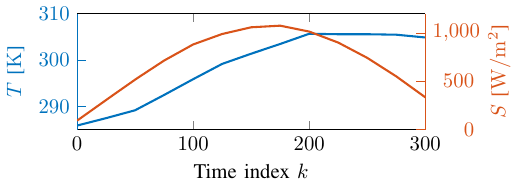}
	\caption{Temperature $T$ (\protect\blueline) and irradiance $I$ (\protect\redline) between 6AM ($k=0$) and 6PM ($k=300$) on a clear and sunny day. \label{fig:temp}}
	\vspace{-20pt}
\end{figure}

Consider a photo-voltaic array with $n_s$ cells in series, with varying temperature $T$ and irradiance $I$ throughout a day as shown in Fig. \ref{fig:temp}. The light-generated current $i_s$ and the reverse saturation current $i_0$ follow from the model in \cite{Vachtsevanos1987}:
\begin{align} \label{eq:is}
	i_s &= \left(I_s + k_i(T-T_r)\right) \textstyle\frac{S}{1000} \\
	i_0 &= I_0 \left(\textstyle\frac{T}{T_r}\right)^3 e^{\frac{E_g}{NV_t} \left(\frac{T}{T_r} -1 \right)}, \quad V_t = \textstyle\frac{kT}{q}, \label{eq:io}
\end{align}
All parameters are given in Table \ref{tab:pars}. The output current $i$ of an array with $n_s$ cells in series is given by
\begin{align}
	i = i_s - i_0\left( e^{\frac{v+iR_s n_s}{N V_t n_s}}-1\right) - \textstyle\frac{v+i R_s n_s}{R_p n_s}. \label{eq:i}
\end{align}
Here, $v$ denotes the output voltage. A dc-dc buck converter is used to connect each array to a dc load. The converter dynamics are represented by the model in \cite{Li2016} as 
\begin{align}
	C_c \dot{v} = i(v;T,S)-i_L u, \quad
	L_c \dot{i}_L = -i_L R_c + vu, \label{eq:doti}
\end{align}
with $i(v;T,S)$ the nonlinear mapping from the duty cycle $u$ to the output current. Using \eqref{eq:is}-\eqref{eq:doti}, the nonlinear steady-state mapping between the duty cycle $u$ and the produced power $P$ can be computed for any combination of $T$ and $I$.

The varying temperature and irradiance in Fig. \ref{fig:temp} lead to variations in the mapping from the duty cycle to the produced power, as illustrated in Fig. \ref{fig:graph}. The aim is to find and track the optimal duty cycle that maximizes the produced power. We assume that the power measurements contain zero-mean Gaussian white noise with standard deviation $\rho = 5$.

\subsection{Results}

\begin{figure}[t]
	\centering
	\begin{subfigure}{\linewidth}
			\setlength\figureheight{.4\figurewidth}
	 \includegraphics{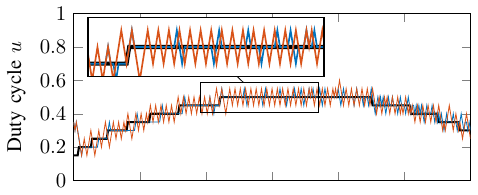}
	\end{subfigure} \\ 
	\begin{subfigure}{\linewidth}
		\setlength\figureheight{.4\figurewidth}
	\includegraphics{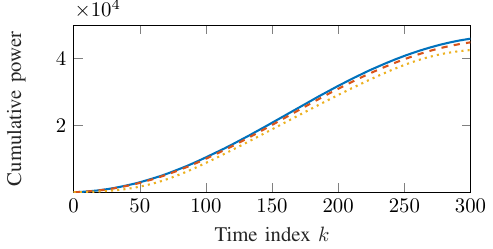}
	\end{subfigure}
	\caption{The reduced number of perturbations (top) in uP\&O (\protect\blueline) leads to a 2.4\% improvement in power output (bottom) over a day compared to P\&O (\protect\reddash), and an 8\% improvement compared to using the optimal constant input $u=0.45$ (\protect\yeldots). \label{fig:cost}}
	\vspace{-20pt}
\end{figure}

Both uP\&O and P\&O are applied to find and track the duty cycle that maximizes the power output. For uP\&O, the design parameters are chosen as $W=0$, $\lambda = 0.88$ and $\hat{\rho} = 5$, i.e., it is assumed that the standard deviation of the measurement noise is known. In Fig. \ref{fig:cost} (top), the inputs selected by the two methods are shown. During the 300 time steps, uP\&O uses 91 perturbations (defined as time instances where the input differs from the optimal input), which is far fewer than the 165 perturbations used by P\&O, yet both methods are capable of tracking the optimal input value as it varies over time. The realized power output for both methods is shown in Fig. \ref{fig:cost} (bottom), showing a 2.4\% increase for uP\&O compared to P\&O, and an 8\% increase compared to using the optimal constant input. This demonstrates that by reducing the number of unnecessary perturbations, uP\&O can increase the produced power significantly.

\section{Conclusions} \label{sec:conclusions}

%
%

In this paper, we present an uncertainty-based perturb-and-observe method. An uncertainty-based model of the cost function is used to determine when to perturb, leading to a vast reduction in the number of perturbations required to accurately track time-varying optima. This leads to improved performance, and may also reduce machinery wear and operating costs, improving the suitability of perturbation-based optimization for industrial applications. Conditions for the convergence of uP\&O are provided. Simulation results from an industrial use case illustrate that the approach achieves fast and accurate tracking with fewer perturbations compared to standard P\&O. Ongoing research involves analyzing the reduction in perturbations, extensions to the multivariable case, the use of model knowledge and experimental validations.




\addtolength{\textheight}{-12cm}   



\bibliography{library.bib}

\end{document}